\documentclass[%
 aip,
nofootinbib,
 amsmath,amssymb,
 reprint,%
]{revtex4-2}

\usepackage{graphicx}
\usepackage{dcolumn}
\usepackage{bm}
\usepackage[utf8]{inputenc}
\usepackage[T1]{fontenc}
\usepackage{mathptmx}
\usepackage{etoolbox}
\usepackage{subcaption}
\usepackage{hyperref}
\usepackage{physics}
\setcitestyle{square,numbers}

\newcommand*{\citen}[1]{%
  \begingroup
    \romannumeral-`\x 
    \cite{#1}%
  \endgroup   
}

\makeatletter
\def\@email#1#2{%
 \endgroup
 \patchcmd{\titleblock@produce}
  {\frontmatter@RRAPformat}
  {\frontmatter@RRAPformat{\produce@RRAP{*#1\href{mailto:#2}{#2}}}\frontmatter@RRAPformat}
  {}{}
}%
\makeatother
\begin{document}

\preprint{AIP/123-QED}

\title[Simulations of the churning mode]{Simulations of the churning mode: toroidally symmetric plasma convection and turbulence around the X-points in a snowflake divertor}
\author{D. Power}
\email{power8@llnl.gov.}
\author{M. V. Umansky}
\author{V. A. Soukhanovskii}
\affiliation{ 
$^1$Lawrence Livermore National Laboratory, Livermore, CA, USA
}

\date{\today}

\begin{abstract}

Using a reduced MHD model, extended to include field-aligned thermal conduction, we present numerical simulations of the churning mode (CM): a toroidally symmetric, non-linear plasma vortex in the vicinity of the null points in a snowflake (SF) divertor (Ryutov et al., Phys. Scr. \textbf{89} 088002, 2014). Simulations are carried out across a range of inter-null separations, $d_{xx}$, and inter-null orientations, $\theta$, primarily in conditions relevant to the MAST-U tokamak. We find that, when $d_{xx}$ is small, the CM induces additional transport across the X-points when $\beta_{pm} \gtrsim 8$ \%, where $\beta_{pm}$ is the ratio of the plasma pressure in the null region to poloidal magnetic pressure at the midplane. This transport also increases approximately linearly as $d_{xx}$ is reduced. A diffusive model of this transport is shown to predict the total transport across the null points, where diffusion coefficients of up to $\sim 10^2$ m$^2$s$^{-1}$ centred on a small region around the X-points are used. However, the CM also results in significant changes to the flux surfaces in the null region which is not captured by this diffusive model. The changes in magnetic geometry mean the fractional exhaust power delivered to each divertor leg is highly sensitive to $\beta_{pm}$, $d_{xx}$ and $\theta$. For small values of $\theta$, the CM can induce a change in topology, redirecting exhaust power from a secondary divertor leg on the high field side to one on the low field side. Similar behaviour is found in the fraction of exhaust power going to the inner and outer divertor. Such changes in the flux surfaces may not be captured by Grad-Shafranov solvers and so may be a source of error in the magnetic reconstruction of SF experiments. We consistently find that the fractional exhaust power going to a secondary divertor leg on the high field side is small, consistent with SF experiments. 

\end{abstract}

\maketitle

\section{\label{sec:intro}Introduction}

In reactor scale tokamak devices, heat and particle loads to the plasma-facing components (PFCs) are predicted to be higher than tolerable material limits. Radiative dissipation of these heat loads in a standard divertor configuration may be insufficient, and so novel approaches have been proposed. One such idea is the snowflake (SF) divertor \cite{ryutov_geometrical_2007,ryutov_snowflake_2015}, which has several favourable exhaust properties. 

A standard single null (SN) divertor features a single X-point. In the SF divertor, a secondary X-point is brought into close proximity with the first one, resulting in a configuration with four legs per divertor instead of the usual two. This doubling of the number of divertor legs, along with an increased connection length, divertor volume and flux expansion \cite{ryutov_geometrical_2007,ryutov_magnetic_2008}, may help to reduce heat and particle loads to PFCs. 

\begin{figure}[h]
  \includegraphics[width=\linewidth]{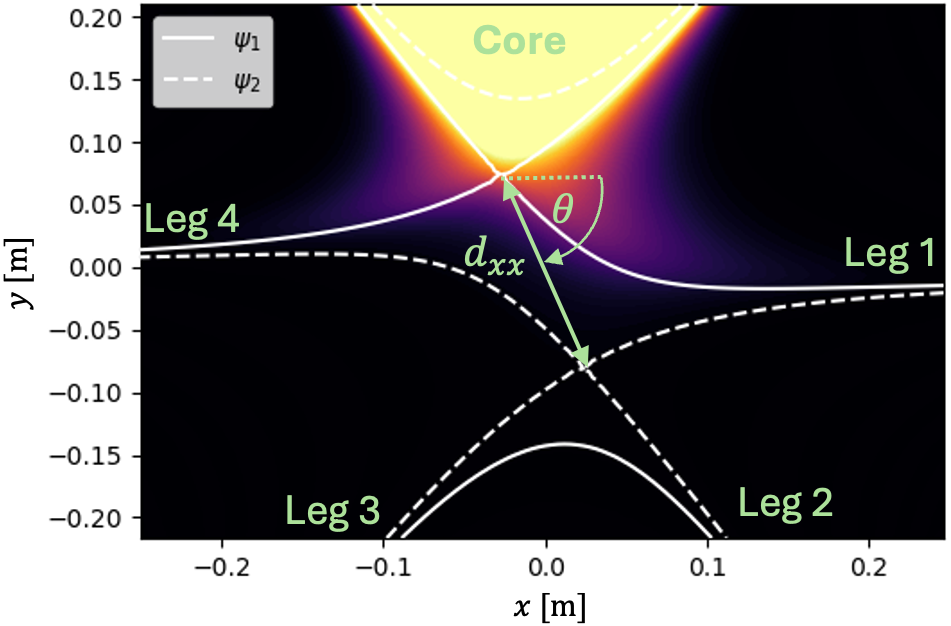}
  \caption{Diagram of the magnetic topology of a SF divertor in a small region close to the X-points. Primary and secondary separatrices at $\psi_1$ and $\psi_2$ are shown in white, $d_{xx}$ is the separation distance between nulls and $\theta$ is the orientation angle. The local magnetic geometry in the null region is determined by $d_{xx}$ and $\theta$. Each divertor leg is assigned a number as shown. The shaded contours show a representative plasma pressure profile for this SF-plus configuration: a hot core, a falling off of pressure in the SOL, and low pressure in the private flux region.}
  \label{fig:schematic}
\end{figure}

In the exact SF, the two X-points are superimposed and there is a null in both the poloidal magnetic field and its derivative at this point. Such a configuration is topologically unstable, so in practice the two X-points are merely brought close together. The separation distance between the nulls $d_{xx}$, and their orientation angle $\theta$, determine the magnetic topology in the vicinity of the null region, see fig. \ref{fig:schematic}. Favourable properties of the SF are maintained in this inexact scenario provided $d_{xx}$ is sufficiently small \cite{ryutov_snowflake_2012}. The last closed flux surface in the core determines the primary X-point and separatrix. When the secondary X-point is in the inner or outer scrape-off layer (SOL), the configuration is called a "snowflake minus" (SF-minus); when it is in the private flux region, it is called a "snowflake plus" (SF-plus).

A key property of the SF is that the poloidal field (PF) strength increases with the square of the distance from the X-points, as opposed to rising linearly in a SN divertor. This results in a larger region of low PF which may facilitate increased cross-field transport \cite{ryutov_snowflake_2012}. Experiments at TCV \cite{piras_snowflake_2010,reimerdes_power_2013,vijvers_power_2014}, DIII-D \cite{soukhanovskii_developing_2018} and MAST-U \cite{Soukhanovskii2022} indicate some evidence of enhanced transport close to the nulls resulting in a redistribution of exhaust power across the four divertor legs. Furthermore, simulations with the edge transport code EMC3-EIRENE have shown \cite{lunt_first_2014} that a region of enhanced cross-field transport around the nulls, beyond that which would be expected from the geometrical properties of the SF divertor alone, is required to match experimental profiles. 


Several mechanisms driving this enhanced transport have been proposed but not yet verified: strong $\vb{E}\times \vb{B}$ drifts \cite{canal_enhanced_2015}, magnetic stochastization \cite{abdullaev_description_2008}, MHD instabilities in the case of pressure gradients anti-parallel to the direction of magnetic curvature \cite{farmer_axisymmetric_2013}, and ballooning modes in the divertor region \cite{farmer_ballooning_2014}. 

A leading candidate, which is the focus of this paper, is the so-called "churning mode" (CM) \cite{ryutov_churning_2014}. Due to the weak PF close to the X-points, it is difficult to satisfy the toroidal equilibrium conditions when there is a vertical pressure gradient (as would be expected at the upper and lower boundaries of the confined core plasma). The result is that toroidally symmetric convective mixing can occur in the zone where the ratio of the plasma pressure to poloidal magnetic pressure is large, $\beta_p \gg 1$. Axisymmetric plasma rotation about the X-points, as described in \cite{ryutov_churning_2014}, may facilitate efficient sharing of the heat and particle flux across some or all of the divertor legs. 

A model of the CM based on reduced MHD was presented in \cite{umansky_toroidally_2016}. There, it was shown that convective motion driven by the interaction bteween toroidal curvature and vertical pressure gradients can result in a 2D (toroidally symmetric) plasma vortex about the X-points in a SF divertor. For a static pressure profile with uniform vertical pressure gradient, parameterised by $\delta$, the size of this convective zone is shown to scale with $\sqrt{ \delta \varepsilon \beta_{pm}}$, where $\varepsilon$ is the ratio of minor radius to major radius of the device, 
\begin{equation}
  \label{eq:epsilon}
  \varepsilon=a/R_0,
\end{equation}
and $\beta_{pm}$ is the poloidal beta measured using the thermal pressure in the null ragion and the magnetic pressure at the outer midplane, 
\begin{equation}
  \beta_{pm}=8\pi p_0/B_{pm}^2.
\end{equation} 
The dependence of the CM on $\varepsilon$ indicates a potentially stronger role in devices with small aspect ratio (large $\varepsilon$, i.e. spherical tokamaks). In addition, the relationship with $\beta_{pm}$ (highlighted also in \cite{ryutov_churning_2014}) suggests that it may play an important role during ELMs. In theory, the higher plasma pressure during ELMs could increase mixing in the divertor region and go some way to mitigating intolerable peak heat fluxes on the target plates. 




The free energy for the CM comes from the vertical pressure gradient sustained by plasma heating in the core and cooling in the divertor. Therefore, thermal transport plays a key role in governing the CM physics. One of the motivations for the study presented here is therefore to extend the analysis in \cite{umansky_toroidally_2016} to include the effect of field-aligned conductive thermal transport. This enables exploration of the long-run behaviour of the mode, where the vertical pressure gradient and convective mixing will jointly determine transport across the null region.

Further motivation comes from the fact that, in \cite{umansky_toroidally_2016}, only the exact SF configuration was considered. This is difficult to obtain experimentally, and may not have ideal exhaust properties anyway. Here, we will explore the role of the CM in more realistic conditions with finite $d_{xx}$ and $\theta$. This will enable a greater understanding of the role of the CM in recent and future SF experiments.  

A final research question we wish to answer is the nature of the CM-driven transport. It is not necessarily desirable or practical to include a full model of the CM in detailed studies of tokamak edge transport for SF experiments. Therefore, there is interest in reduced models which capture the dominant physics in a more analytically or numerically tractable manner. In \cite{khrabry_modeling_2021} a diffusive model was proposed, and the simulations carried out here will enable validation of this model and its application to non-exact SFs. 

This study will focus on plasma conditions relevant to the MAST-U device, a medium-sized spherical tokamak with minor radius $a=65$ cm and major radius $R_0=85$ cm. A key focus of MAST-U research is to explore novel divertor concepts, and SF experiments have recently been carried out on this device \cite{Soukhanovskii2022}. Studies of edge transport in MAST-U conditions with a SF divertor are therefore relevant to both recent experiments and future planned research. 

The paper is structured as follows: the model presented in \cite{umansky_toroidally_2016} is extended to include thermal transport in sec. \ref{sec:model}. Numerical simulations relevant to the MAST-U tokamak have been carried out, and these are described in sec. \ref{sec:simulations}. Results are presented in sec. \ref{sec:results}, followed by a scaling of the observed CM transport to different devices in sec. \ref{sec:scaling}. The viability of reduced models of the CM is also discussed there. There is a discussion of the results in in sec. \ref{sec:conclusions}, followed by a brief summary of the findings in sec. \ref{sec:conclusions}. There is also an appendix describing the numerical implementation of field-aligned thermal transport in the model, Appendix \ref{app:cond}.

\section{Churning mode model}
\label{sec:model}


A model of the churning mode has been presented by Umansky \& Ryutov in \citen{umansky_toroidally_2016}. This model is adopted here, extended to include field-aligned thermal conduction. 
This addition therefore allows us to explore the interplay between convective and conductive transport in CM turbulence. 

We work in the same geometry as \cite{umansky_toroidally_2016}: Cartesian $x$-$y$ coordinates in the poloidal plane, centred at $(R_0,Z_0)$ such that $x=R-R_0$ and $y=Z-Z_0$, where $R$ and $Z$ are the radial and vertical coordinates in cyclindrical geometry. The origin is chosen so that one or both null points lie at the point $(R_0,Z_0)$. The model equations are
\begin{equation}
  \label{eq:orig_vort}
      \frac{d \varpi}{d t} =-\frac{2}{R_0} \pderivative{p}{y} +\frac{1}{4 \pi R_0^2}\left\{\psi, \grad^2 \psi\right\},
\end{equation}
\begin{equation}
  \label{eq:orig_psi}
  \frac{d \psi}{d t} =0,
\end{equation}
and
\begin{equation}
  \label{eq:orig_p}
  \frac{3}{2}\frac{dp}{d t}= -\div{\bm{q}},
\end{equation}
where $\varpi=\frac{c}{B_{t0}}\rho \grad^2 \Phi$ is the plasma vorticity, $c$ is the speed of light, $B_{t0}$ is the toroidal magnetic field strength on axis, $\rho=m_i n$ is the plasma mass density, $\Phi$ is the electric potential, $p$ is the total (isotropic) plasma pressure, 
$\psi$ is the poloidal magnetic flux function and $R_0$ is the major radius.
The bracket is
\begin{equation}
  \{f,g\} = \frac{\partial f}{\partial x} \frac{\partial g}{\partial y} - \frac{\partial g}{\partial x} \frac{\partial f}{\partial y},
\end{equation}
the Laplacian is 
\begin{equation}
  \grad^2 = \pdv[2]{x} + \pdv[2]{y},
\end{equation}
and the convective derivative is
\begin{equation}
  \frac{df}{dt} = \frac{\partial f}{\partial t} + \bm{v} \cdot \grad f=\frac{\partial f}{\partial t} + \frac{c}{B_{t0}}\{f,\Phi\},
\end{equation}
where $\bm{v}$ is the fluid velocity. The conductive heat flux is 
\begin{equation}
  \vb{q} = \chi_\parallel \grad_\parallel{T_e} + \chi_\perp \grad_\perp{T_e},
\end{equation}
where $T_e$ is the electron temperature, $T_e=T_i=p/2n$, and $\chi_\parallel$ and $\chi_\perp$ are the parallel and perpendicular thermal conductivities.  
The parallel and perpendicular derivates are $\grad_\parallel = \vb{b}(\vb{b}\cdot\grad)$, $\grad_\perp = \grad - \grad_\parallel$, where $\bm{b}=\vb{B}/B$ is the unit vector parallel to the magnetic field, $\vb{B}=\frac{1}{x+R_0}(B_{t0}\hat{z} + \hat{z} \times \grad{\psi})$. 

This model is similar to reduced MHD models provided in the literature, e.g. \cite{hazeltine_shear-alfven_1985,drake_nonlinear_1984}, and is derived from ideal MHD based on the same expansion in the small parameter $\varepsilon$, here taken to be the inverse aspect ratio (eq. \ref{eq:epsilon}) with the minor radius measured at the outer midplane. For application to studies of the churning mode, we assume constant mass density, incompressible flow, toroidal symmetry, and we have used the lowest-order approximation of the magnetic curvature in the vorticity equation (eq. \ref{eq:orig_vort}), $\vb{\kappa}\simeq -\frac{1}{R_0} \hat{x}$. This model enables studies of MHD turbulence in the vicinity of the X-point(s).

The model in eqs. \ref{eq:orig_vort} - \ref{eq:orig_p} describes two competing processes in the convective transport. The first term on the left of eq. \ref{eq:orig_vort} is a curvature drive, where vertical pressure gradients can induce a toroidally symmetric, clockwise rotation in the poloidal plane. This is similar to the free thermal convection of a neutral fluid in a vertical gravity field heated from the side, as described in \cite{bratsun_non-linear_2003}. The second term is the magnetic restoring force, resisting bending of the field lines. The equations for $\psi$ and $p$ just describe passive advection in the poloidal plane and (for $p$) conductive thermal transport. In the region relevant to the CM, this thermal transport sustains the vertical pressure gradient. Therefore this model contains sufficient physics to explore the interplay between the dominant turbulent and collisional transport processes in the CM. 

\section{Implementation and simulation details}
\label{sec:simulations}

The model in eqs. \ref{eq:orig_vort} - \ref{eq:orig_p} has been implemented in BOUT++ \cite{Dudson2009}, a framework for numerical simulation of tokamak boundary turbulence and transport models. Spatial derivatives are calculated using finite difference methods, and time integration is carried out using the CVODE implicit solver \cite{hindmarsh_sundials_2005}. 

For the $\div{\vb{q}}$ term in eq. \ref{eq:orig_p}, strong anisotropy in transport parallel and perpendicular to the magnetic field means a robust numerical approach is required. For this we use a modified version of the field line map approach described by Stegmeir et al. in \cite{stegmeir_field_2016}, adapted to our 2D problem domain in which the poloidal magnetic field can point in any direction relative to the grid. See Appendix \ref{app:cond} for details. 

A uniform, rectangular 2D grid in the $x$ and $y$ directions is used, shown in fig. \ref{fig:schematic}, with $n_x$, $n_y$ cells in each direction. The domain size $L_x$ is specified at input, and grid widths are $\Delta x= L_x/n_x$, $\Delta y = \sqrt{3} \Delta x$. This choice of $\Delta y$ is motivated by the fact that the dominant transport directions in a snowflake configuration are $\varphi=n\pi/3$, where $\varphi$ is the angle from from the $x$-axis and $n=0-5$. Aligning the grid with these directions minimises numerical errors in the parallel heat flux calculation \cite{umansky_numerical_2005}.

For the boundary conditions, we set $\Phi=0$ and $\pdv{t}=0$ for $\varpi$, $\psi$ and $p$. The pressure is set to a small value corresponding to $T_e=10$ eV on all boundaries apart from within the `core', defined as the region where $\psi>\psi_1$ on the upper boundary, where $\psi_1$ is the poloidal flux on the primary separatrix. In this region $p$ is fixed to its initial value (discussed shortly). Outside this region on the upper boundary we set $\vb{q}=0$, ensuring that thermal energy in the simulation domain primarily enters via the core, and leaves via the downstream boundaries (left, lower and right edges of the 2D domain). 

All simulations are initialised with $\varpi=p=0$. On the upper boundary, we set pressure to be a function of the normalised flux reducing to zero at the primary separatrix, $p=\delta \sqrt{1-\psi_n}$, where $\psi_n=\frac{\psi - \psi_a}{\psi_1 - \psi_a}$, $\psi_a$ is the poloidal flux on the magnetic axis and $\delta$ is an input parameter. The pressure is allowed to reach equlibrium (solving eq. \ref{eq:orig_p} only) before the full CM model is evolved (eqs. \ref{eq:orig_vort} - \ref{eq:orig_p}). This allows us to replicate conditions similar to those obtained by edge codes such as UEDGE \cite{rognlien_fully_1992} before simulating the effect of the CM physics. 

As discussed, $d_{xx}$ and $\theta$ determine the local magnetic geometry of the null region (see fig. \ref{fig:schematic}). An expression for $\psi$ involving just these parameters in the vicinity of the nulls in a SF divertor has been presented in \cite{ryutov_local_2010}. A similar form is used here to initialise $\psi$ in the simulations, modified to enable cross-comparison of SFs at different values of $d_{xx}$, 
\begin{equation}
  \label{eq:psi_init}
  \psi /A = -\frac{3}{2} \tilde{x}^2\sin{\theta} + 3\tilde{x}\tilde{y}\cos{\theta}+ \frac{3}{2}\tilde{y}^2\sin{\theta} -3\tilde{x}^2\tilde{y}+\tilde{y}^3
\end{equation}
where $A$ sets the scale of the poloidal flux, $\tilde{x}=x/d_{xx}$ and $\tilde{y}=y/d_{xx}$. The values of $\theta$ and $d_{xx}$ are input parameters and $A=A_{0}\left(\frac{d_{xx,0}}{d_{xx}}\right)^3$, where $A_0$ is calculated by fitting eq. \ref{eq:psi_init} in the vicinity of the nulls to a reference magnetic equilibrium from a Grad-Shafranov code with separation $d_{xx,0}$. 

All simulations are for a nominally representative MAST-U SF divertor, with minor radius $a=65$ cm, major radius $R_0=85$ cm, thus $\varepsilon=0.76$. $A_0$ is fitted to a reference MAST-U magnetic equilibrium (modelled with FIESTA) with $d_{xx,0}=2.5$ cm and $B_{pm}=0.3$ T. The toroidal field strength on axis is $B_{t0}=0.8$ T. The values of $p_0$, $d_{xx}$ and $\theta$ will be varied. The parallel thermal conductivity $\chi_\parallel$ is set to a constant equal to the Spitzer-Härm value at the electron temperature in the null region, $T_{sepx}\simeq100$ eV\footnote{Self-consistent Spitzer-Härm conduction has been implemented, but is numerically unstable in the hot core region of the simulation domain. We aim to fix this limitation in future work.}. The SOL width at the outer midplane, $\lambda_{mp}$, does not directly enter the model but can be approximately related to the width at the null region, $\lambda_x$. The value of $\lambda_x$ is tuned in the simulations by varying $\chi_\perp$ (typically $\chi_\perp \sim 10^{-1} - 10^1$ m$^2$s$^{-1}$) and the size of the simulation domain, and then $\lambda_x$ can be related to $\lambda_{mp}$ via the geometric expansion,
\begin{equation}
  \label{eq:d_sol}
  \lambda_x\simeq a (\lambda_{mp}/a)^{1/3}.
\end{equation}

Note that all simulation parameters quoted from here onwards ($p_0$, $\beta_{pm}$, $\theta$, $d_{xx}$ and $\lambda_x$) are the values at the point when heat conduction (eq. \ref{eq:orig_p}) has reached equilibrium and before the full CM model is evolved. 


\section{Simulation results}
\label{sec:results}

\subsection{Churning mode onset}


\begin{figure}[h]
  \includegraphics[width=\linewidth]{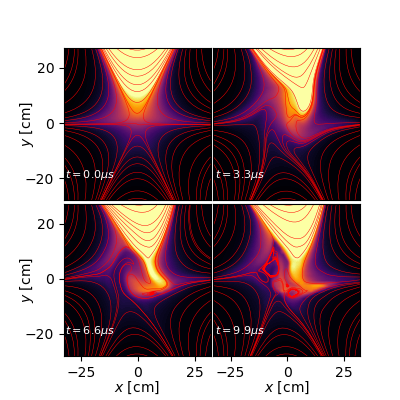}
  \caption{Various time slices after the CM physics is turned on for a thermally equlibrated plasma pressure profile for a representative MAST-U SF divertor. Poloidal magnetic field lines are shown in red.}
  \label{fig:churn_onset}
\end{figure}

Figure \ref{fig:churn_onset} shows the onset of the churning mode in a thermally equilbrated plasma in a near-exact SF configuration, $d_{xx}=2.5$ cm and $\beta_{pm}=0.12$. As in \cite{umansky_toroidally_2016}, clockwise rotation of the plasma occurs initially in a small region around the X-points. Turbulent mixing in this region follows at later times. 

\subsection{Proximity condition: scan in $d_{xx}$}

For an inexact SF with finite $d_{xx}$, there is expected to be a point at which it will resemble an exact SF in terms of exhaust properties as $d_{xx}$ is reduced. One such `proximity condition', based purely on geometrical considerations, is presented in \cite{ryutov_snowflake_2015} and relates $d_{xx}$ to the projection of the SOL width to the null region,
\begin{equation}
  \label{eq:prox_cond}
  d_{xx} < \lambda_x,
\end{equation}
where $\lambda_x$ is defined in eq. \ref{eq:d_sol}. Another proximity condition with relevance to the CM can be formulated, which says the inexact SF will undergo convective mixing as in the exact SF case when 
\begin{equation}
  \label{eq:prox_cond2}
  d_{xx} < r_{cz},
\end{equation}
where
\begin{equation}
  r_{cz}\simeq 0.81 a (\beta_{pm} \varepsilon)^{1/3}
  \label{eq:r_cz}
\end{equation}
is the predicted radius of the convective zone \cite{ryutov_churning_2014}.

\begin{figure}
  \includegraphics[width=\linewidth]{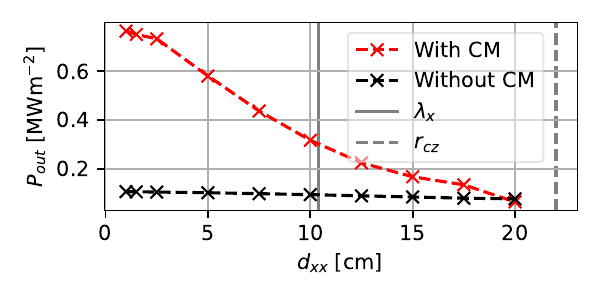}
  \caption{Total output power with and without the CM physics as the inter-null separation distance is varied.}
  \label{fig:dxx_scan}
\end{figure}

The degree to which these conditions hold for CM transport has been assessed by varying $d_{xx}$ from 1.0 to 20.0 cm at an approximately constant $\beta_{pm}\simeq0.10$. The SOL width is $\lambda_x=10.4$ cm, corresponding to $\lambda_{mp}\simeq 2.5$ mm, and the null orientation is $\theta=90^{\circ}$ (i.e. an exact SF-plus configuration). The model equations are evolved until saturation is reached after $t\sim 3$ ms, and output quantities are time-averaged. We plot the total power out of the downstream boundaries
\begin{equation}
  P_{out}=\int \bm{q} \cdot \hat{n} dl,
\end{equation}
where $\hat{n}$ points out of the simulation domain and the integration is over the left, lower and right edges. Given $p$ is fixed on the boundaries, higher $P_{out}$ corresponds to increased transport across the null region. Results are shown in Figure \ref{fig:dxx_scan}: we see that $P_{out}$ increases steadily as $d_{xx}$ is decreased, reaching a saturation point around $d_{xx}=2.5$ cm. This suggests that, for CM transport, a more stringent proximity condition than eq. \ref{eq:prox_cond} or \ref{eq:prox_cond2} applies. 

\subsection{Activation of CM transport: scan in $\beta_{pm}$}
\begin{figure}
  \includegraphics[width=0.9\linewidth]{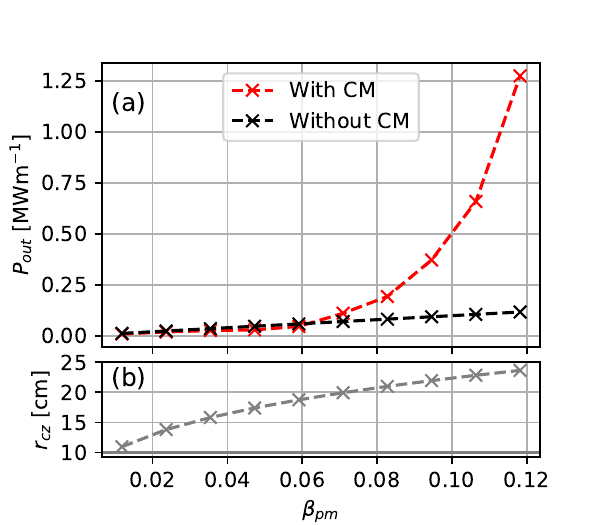}
  \caption{(a): Total output power in a density scan with and without the CM physics, plotted as a function of the normalised poloidal beta. (b): Predicted radius of the convective zone.}
  \label{fig:nscan}
\end{figure}

To observe the point at which the churning mode `activates' in the near-exact SF case as pressure is increased, we vary the plasma density (which is fixed in our model) from $n=10^{19}-10^{20}$ m$^{-3}$, resulting in $\beta_{pm}=0.01-0.12$. The SOL width is again $\lambda_x=10.4$ cm. Results are shown in fig. \ref{fig:nscan}a, plotted as a function of $\beta_{pm}$. We see that the CM is driving additional transport at $\beta_{pm}\gtrsim 0.08$, increasing rapidly with $\beta_{pm}$ and resulting in a roughly order of magnitude increase in $P_{out}$ at the highest $\beta_{pm}$. 

We might expect that the CM induces significant transport across the null region when $r_{cz}\gtrsim \lambda_x$, i.e. when the majority of the SOL is subject to convective transport. Fig. \ref{fig:nscan}b shows the predicted values of $r_{cz}$ using eq. \ref{eq:r_cz} in these simulations: noting that $\lambda_x=10.4$ cm here, we see that significant transport across the nulls occurs for $r_{cz} \gtrsim 2 \lambda_x$.

\subsection{Divertor leg power sharing}

\begin{figure}
  \begin{subfigure}[h]{\linewidth}
    \includegraphics[width=\linewidth]{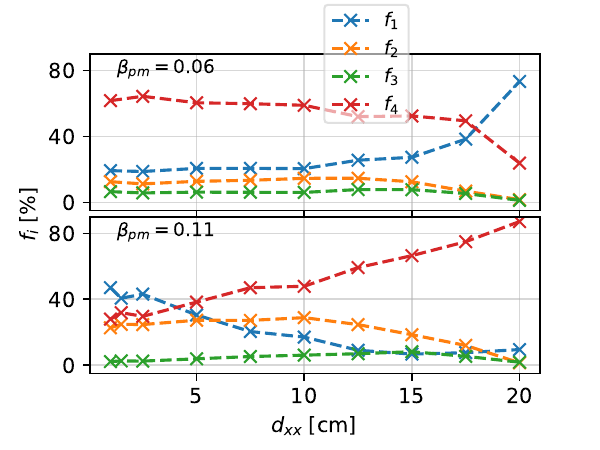}
    \caption{}
  \end{subfigure}
  \begin{subfigure}[h]{\linewidth}
    \includegraphics[width=\linewidth]{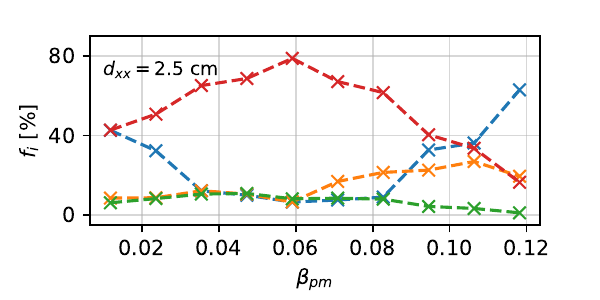}
    \caption{}
  \end{subfigure}
  \caption{Fractional output power going to each divertor leg: (a) in a scan of $d_{xx}$ at two values of $\beta_{pm}$; (b) in a scan of $\beta_{pm}$ at $d_{xx}=2.5$ cm.}
  \label{fig:Q_legs}
\end{figure}

In SF experiments, evidence of enhanced transport across the null region has been shown in the form of an increase in exhaust power and particle flux to secondary strike points \cite{vijvers_power_2014,soukhanovskii_developing_2018}.  In fig. \ref{fig:Q_legs}, we show the fractional power to each divertor leg from the simulations,
\begin{equation}
  f_i = P_i/P_{out},
\end{equation}
where $i=1,2,3,4$ and $\sum_i P_i = P_{out}$. This quantity is shown (a) in $d_{xx}$ scans at two values of $\beta_{pm}$; (b) in a $\beta_{pm}$ scan at fixed $d_{xx}=2.5$ cm. For the SF-plus configuration simulated here ($\theta=90^\circ$), the secondary divertor legs are legs 2 and 3. The first thing to note is that the CM does result in some activation of the secondary legs as $d_{xx}$ is reduced from 20 cm. But when the turbulence is strong at higher $\beta_{pm}$, power is only effectively shared across three legs at low $d_{xx}$.

\begin{figure}
  \includegraphics[width=\linewidth]{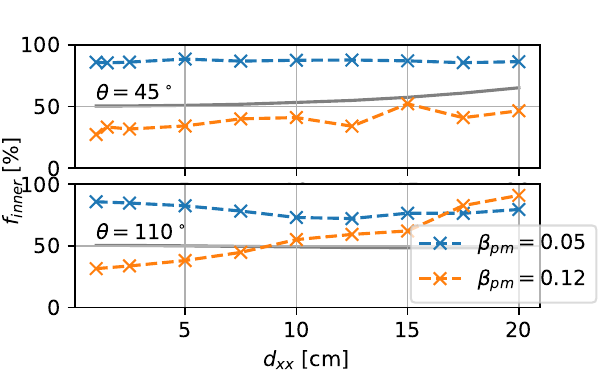}
  \caption{Fractional power delivered to the inner divertor in scans in $d_{xx}$ at two values of $\theta$ and two values of $\beta_{pm}$. Values of $f_{inner}$ without the CM are shown in solid grey.}
  \label{fig:f_inner_theta}
\end{figure}

While the CM can increase exhaust power to secondary divertor legs, it is also useful to analyse the fractional power delivered to the inner or outer divertor as this can be easier to measure experimentally. In fig. \ref{fig:f_inner_theta} we plot the exhaust power going to the inner divertor legs,
\begin{equation}
  f_{inner}=(P_3 + P_4)/P_{out},
\end{equation}
in $d_{xx}$ scans at two values of $\beta_{pm}$ and two values of $\theta$. This shows that the CM can strongly influence $f_{inner}$. At lower $\beta_{pm}$, $f_{inner}$ is roughly constant in $d_{xx}$ but still modified from the case without the CM. At high $\beta_{pm}$, $f_{inner}$ is more sensitive to $d_{xx}$. 

\subsection{Validation of a diffusive model of CM transport}
In \cite{khrabry_modeling_2021,khrabry_modeling_2022}, a diffusive `umbrella' model of transport in the null region is proposed and used in SF divertor modelling of MAST-U in UEDGE. Two Gaussian profiles of enhanced perpendicular diffusion are added to $\chi_\perp$, one centred on each null point, with a size given by eq. \ref{eq:r_cz}. The maximum theoretical value of the diffusion coefficient is estimated as \cite{ryutov_churning_2014} 
\begin{equation}
  \chi_{x}\simeq \frac{1}{2}r_{cz}^2 / \tau,
  \label{eq:pred_chi_x}
\end{equation}
where $\tau=\simeq \sqrt{R_0 r_{cz}}/v_{th,i}$ is a predicted turnover time for the churning mode and $v_{th,i}$ is the ion thermal velocity at $p_0$. Thus the perpendicular diffusion coefficients are given by 
\begin{equation}
  \chi_{\perp,mix} = \chi_\perp + \chi_{add},
  \label{eq:chi_mix}
\end{equation}
where 
\begin{equation}
  \begin{aligned}
    \chi_{add} = \chi_x & \bigg[e^{-\left(\frac{x-x_1}{r_{cz}}\right)^2 -\left(\frac{y-y_1}{r_{cz}}\right)^2}  \\
    & + e^{-\left(\frac{x-x_2}{r_{cz}}\right)^2 -\left(\frac{y-y_2}{r_{cz}}\right)^2} \bigg],
  \end{aligned}
  \label{eq:chi_x}
\end{equation}
and $(x_i,y_i)$ are the coordinates of each null point. 

\begin{figure}
  \includegraphics[width=\linewidth]{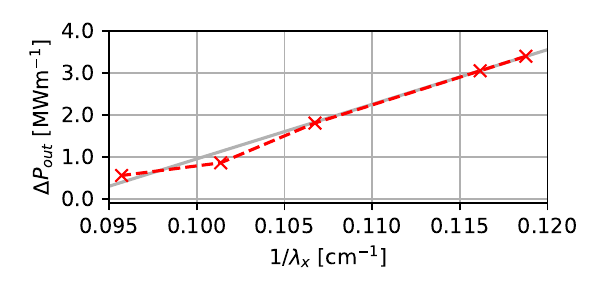}
  \caption{Increase in $P_{out}$ due to the churning mode vs. $1/\lambda_x$ at fixed $\beta_{pm}$. An approximate linear fit is shown in grey.}
  \label{fig:dsolscan}
\end{figure}

To determine the validity of this approach, first we assess whether CM transport can be approximated diffusively. This is done by varying the parameter $\lambda_x$: a diffusive perpendicular transport model would predict $P_{out}\propto \grad_\perp p$, where $\grad_\perp p \sim p_0/\lambda_x$ in the null region. Thus if $P_{out} \propto 1/\lambda_x$, this indicates a diffusive model is appropriate. To test this, $\lambda_x$ is varied from 9.4 cm to 12.1 cm ($\lambda_{mp}=1.3-2.7$ mm) at $\beta_{pm}=0.11$. These parameters are chosen to ensure $r_{cz}\gg \lambda_x$ and therefore that the CM is `active'. $\Delta P_{out}$ is the difference in $P_{out}$ with/without the CM physics, and this is plotted in fig. \ref{fig:dsolscan}. We see that $\Delta P_{out}$ does indeed scale approximately linearly with $1/\lambda_x$ here. At high $\lambda_x$ (low $1/\lambda_x$), there is some deviation from the linear trend as $r_{cz}\gg \lambda_x$ is not adequately satisfied and $\Delta P_{out}$ approaches zero.

\begin{figure}
  \includegraphics[width=\linewidth]{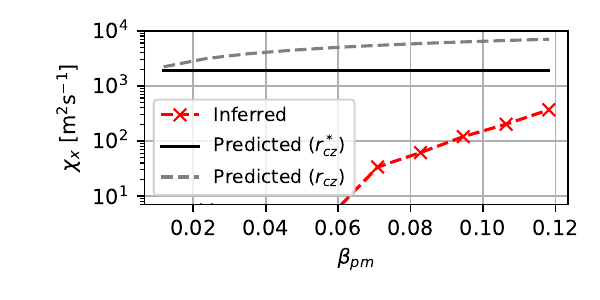}
  \caption{Estimated values of the peak diffusive coefficient in the umbrella model used in \cite{khrabry_modeling_2021,khrabry_modeling_2022} from CM simulations. Maximum theoretical values using an expression from \cite{ryutov_churning_2014} are shown in red.}
  \label{fig:chi_x}
\end{figure}

We will now make rough estimates of $\chi_x$ in eq. \ref{eq:chi_x} for this diffusive umbrella model from calculations of $P_{out}$ in the simulations. Let $\chi_\perp^{eff}$ be an effective diffusion coefficient across the null region, then we have an effective heat flux given by
\begin{equation}
  \vb{q}_\perp^{eff} = \chi_\perp^{eff}\grad_\perp{p}.
\end{equation}
From the definition of $P_{out}$ we have $\abs{\vb{q}_\perp^{eff}} = P_{out} / L_{d}$, where $L_{d}$ is the length of the downstream boundaries of the simulation domain. Using $\grad_\perp{p}\simeq p_0/\lambda_x$ and letting the volume average of $\chi_{add}$ be $\langle \chi_{add} \rangle=\chi_\perp^{eff} - \chi_\perp$, we get
\begin{equation}
  \chi_x \simeq \left( \frac{P_{out}\lambda_x}{p_0 L_{d}} - \chi_\perp \right) \frac{2A}{\pi (r_{cz}^*)^2},
  \label{eq:chi_x_estimate}
\end{equation}
where we have integrated over eq. \ref{eq:chi_x} for the exact SF case and $A$ is the size of the simulation domain. We have also used $r_{cz}^*=\min(r_{cz},\lambda_x)$ as the radius of the diffusive zone, assuming it cannot be greater than the SOL width in the null region. 
Inferred values of $\chi_x$ for the density scan from fig. \ref{fig:nscan} are shown in fig. \ref{fig:chi_x}, alongside predicted maximum values using eq. \ref{eq:pred_chi_x} using $r_{cz}$ (dashed grey) and $r_{cz}^*$ (solid black). After activation of the CM ($\beta_{pm} \gtrsim 0.08$), where $\chi_x$ is small, inferred values are still smaller than predicted but are increasing rapidly. As such, eq. \ref{eq:pred_chi_x} may be applicable only in the limit of high $\beta_{pm}$. 


\subsection{Distortion of flux surfaces}

\begin{figure}
  \includegraphics[width=\linewidth]{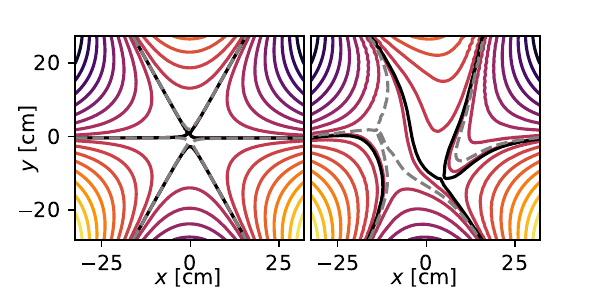}
  \caption{Contours of the poloidal flux in a CM simulation with $d_{xx}=2.5$ cm and $\beta_{pm}=0.10$ at initialisation. Primary and secondary separatrices are shown in black and grey respectively. Left: at $t=0$. Right: time-averaged contours after $t\sim 3$ ms.}
  \label{fig:distorted_psi}
\end{figure}

Another possible explanation for the smaller-than-predicted values of $\chi_x$ (fig. \ref{fig:chi_x}) is related to a rearrangement of the flux surfaces seen in the CM simulations: 
in the exact SF case ($d_{xx}\sim 0$) the CM forces a reconfiguration resulting in an inexact SF, where smaller cross-field transport would be predicted. 
In fig. \ref{fig:distorted_psi}, we plot contours of the poloidal flux $\psi$ at initialisation and after a quasi-equilibrium state has been reached at $t\sim 3$ ms for the $\beta_{pm}=0.10$ case from the density scan in fig. \ref{fig:nscan}. Contours at the later time are time-averaged. 
We see that the position of the separatrices in flux coordinates has moved (discussed shortly) and $d_{xx}$ is now around 20 cm. This change in geometry is reflected in the distribution of power to each divertor leg, $f_i$. Values of $f_i$ are shown in table \ref{tbl:power_fracs}: rather than an approximately even split across all four legs, leg 2 receives almost ten times more power than leg 3. Note that this situation is replicated at different values of $\theta$: at such small $d_{xx}$, the null orientation does not affect quasi-equilibrium magnetic geometry or exhaust power distribution.  

\begin{table}[h]
  \centering
  \begin{tabular}{ l|l|l|l|l } 
     &$f_1$ & $f_2$ & $f_3$ & $f_4$ \\ 
    \hline
    $t=0$ & 47\% & 2\% & 2\% & 48\% \\
    $t\sim 3$ ms & 35\% & 27\% & 3\% & 35\% 
  \end{tabular}
  \caption{Fraction of divertor power going to each leg in the simulation in fig. \ref{fig:distorted_psi}. See fig. \ref{fig:schematic} for leg numbering. Percentages do not add up to 100 due to rounding. }
  \label{tbl:power_fracs}
\end{table}

\begin{figure}
  \includegraphics[width=\linewidth]{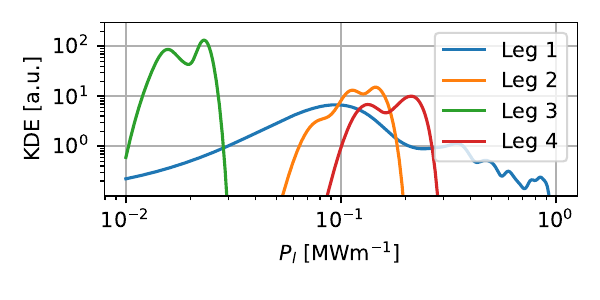}
  \caption{Spread in exhaust power delivered to each divertor leg after quasi-equilibrium state has been reached.}
  \label{fig:P_l_kdes}
\end{figure}

Fig. \ref{fig:distorted_psi} and table \ref{tbl:power_fracs} show the time-averaged picture, but the CM model does not reach a true equilibrium state and fluctuations remain. The spread of values of the power deposited to each divertor leg, $P_i$ where $l=1,2,3,4$ and $\sum_i P_i = P_{out}$, is shown via kernel density estimation in fig. \ref{fig:P_l_kdes}. $P_i$ is calculated for each leg at each timestep after the quasi-equilibrium state has been reached at $t_{eq}=2.9$ ms. The spread in $P_1$ is large, but leg 3 persistently receives very little power, consistent with the magnetic geometry in fig. \ref{fig:distorted_psi} (right panel).

\subsection{Separatrix fluctuations}

\begin{figure}
  \includegraphics[width=\linewidth]{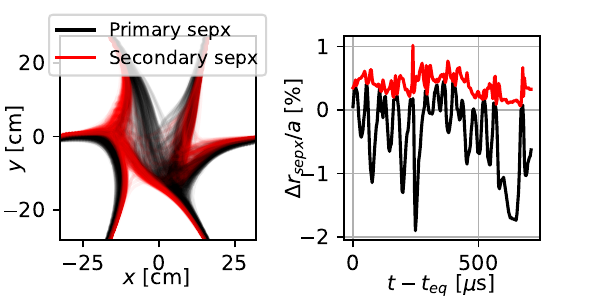}
  \caption{Change in separatrix positions at each timestep after quasi-equilibrium reached at $t_{eq}=2.9$ ms. Left: overlaid contours of $\psi_1$, $\psi_2$. Right: Change in radial position of the separatrices over time, projected to the outer midplane.}
  \label{fig:sepx_dance}
\end{figure}

Fluctuations also occur in the separatrices in these simulations, which may be expected to affect the confinement performance of the plasma. To assess this behaviour in the simulations, we calculate the position of both X-points in flux coordinates at each timestep after quasi-equilibrium in the CM turbulence has been reached. Then we project this location to the outer midplane and convert to a radial coordinate $r$, and compute the difference with the separatrix positions at initialisation. This quantity, $\Delta r_{sepx}$, is shown as a proportion of the minor radius in fig. \ref{fig:sepx_dance}. Fluctuations in the position of the secondary separatrix are small, while for the primary separatrix they are on the order of 1-2\% of $a$ ($\sim 1$ cm), occuring on timescales of $\sim 20\ \mu$s (somewhat larger than the predicted CM turnover time \cite{ryutov_churning_2014} of $\tau\simeq 4\ \mu$s here). While the magnitude of these fluctuations represents a potential loss of a few percent of the confined plasma volume, the diffusive perpendicular transport timescale over this distance for a H-mode plasma with $\chi_\perp= 0.1$ m$^2$s$^{-1}$ is much longer at $\sim 200\ \mu $s. Therefore, the CM-induced fluctuations in the position of the primary separatrix are likely to occur too fast to significantly impact energy and particle losses to the SOL. Note that, in experimental SF studies, no loss of core confinement performance is observed \cite{piras_snowflake_2010,soukhanovskii_taming_2011,soukhanovskii_developing_2018}.

\subsection{Change in magnetic topology}

\begin{figure*}
  \includegraphics[width=\textwidth]{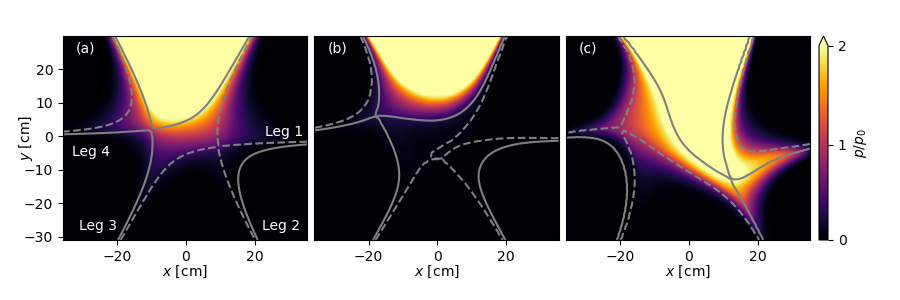}
  \caption{Change in magnetic topology due to the CM above a threshold $\beta_{pm}$. Pressure profiles are shaded, the primary (solid) and secondary (dashed) separatrices are shown in grey. (a): At initialisation, no CM. (b): With CM after quasi-equilibrium reached, low $\beta_{pm}$. (c): With CM after quasi-equilibrium reached, high $\beta_{pm}$. Note that the colour scale for pressure is clipped at $p=2p_0$. 
  }
  \label{fig:topo_change}
\end{figure*}

A final interesting result from the simulations is a CM-induced change in magnetic topology under certain conditions. It is known that the SF-minus configuration, when $\theta=0$, is topologically unstable: small changes in plasma current can change where the secondary X-point lies (inner SOL, outer SOL or private flux region), and therefore how exhaust power is directed towards the target plates. Similar behaviour is observed in the simulations here due to the CM when $\theta$ is merely small and $d_{xx}$ is large, $d_{xx} > \lambda_x$. To demonstrate this, a density scan was carried out resulting in a range of $\beta_{pm}$ from 0.02 to 0.25, where the nulls are separated by $\theta=15^\circ$ and $d_{xx}=20$ cm. The SOL width at the null region is $\lambda_x=10$ cm. The three panels of fig. \ref{fig:topo_change} demonstrate this topology change: 
\begin{enumerate}
  \item[(a):] at initialisation, the secondary X-point lies in the outer SOL,
  \item[(b):] after quasi-equilibrium reached with the CM physics at low $\beta_{pm}$, the magnetic geometry in the null region is distorted but topologically unchanged,
  \item[(c):] at higher $\beta_{pm}$ with the CM physics, the configuration is effectively mirrored: the secondary X-point lies in the inner SOL.
\end{enumerate}

\begin{figure}
  \includegraphics[width=\linewidth]{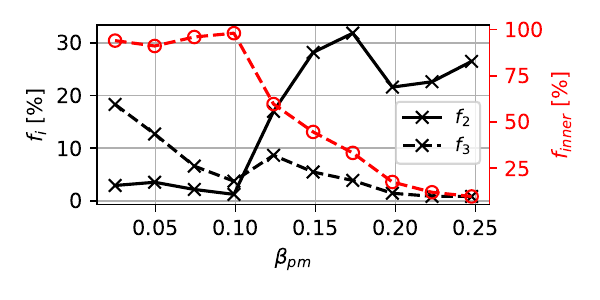}
  \caption{Fraction of the total exhaust power leaving the simulation domain via divertor leg 2 (solid black), leg 3 (dashed black) and via the inner legs (dashed red) for a scan in $\beta_{pm}$ with $\theta=15^\circ$ and $d_{xx}=20$ cm.}
  \label{fig:P2_P3}
\end{figure}

Such a change in topology would redirect exhaust power from divertor leg 3 (lower left) to leg 2 (lower right).
Fig. \ref{fig:P2_P3} shows the fraction of $P_{out}$ leaving via each of these legs, $f_2$ and $f_3$ where $f_i=P_i/P_{out}$, showing that this is indeed the case: the ratio $f_2 / f_3$ goes from close to zero at low $\beta_{pm}$ to around 30 at high $\beta_{pm}$. There is also a significant redirection of power from the inner to the outer divertor legs, where the quantity $f_{inner}=(P_3 + P_4)/P_{out}$ is shown in red.

\section{Transport scaling and assessment of reduced models of the CM}
\label{sec:scaling}

A useful output of this work would be the development of a reduced model which captures the main features of the CM turbulence in a fluid transport model (such as the models used in UEDGE or SOLPS). In fig. \ref{fig:chi_x} we have estimated peak diffusion coefficients for thermal transport across the null region due to the CM turbulence when $d_{xx}$ is small, using the diffusive umbrella model presented in \cite{khrabry_modeling_2021}. 
We can empirically estimate a scaling law for $\chi_x$ by simulating conditions representative of several devices which have carried out SF divertor experiments, shown in table \ref{tbl:device_conditions}, for $d_{xx}$ close to zero. By varying the null pressure and $\chi_\perp$ we can simulate a range of $\beta_{pm}$ and $\lambda_{mp}$. The result is that we find 
\begin{equation}
  \chi_x \simeq 1.90 \varepsilon^{-3.10} \beta_{pm}^{2.90} \lambda_{mp}^{-1.74} 
  \label{eq:chi_x_fit_2}
\end{equation}
fits the inferred values of $\chi_x$ well with a root mean square error of 8\%. 

\begin{table}[h]
  \centering
  \begin{tabular}{ l|l|l|l } 
    Device &$a$ [m] & $R_0$ [m] & $B_{pm}$ [T] \\ 
    \hline
    MAST-U & 0.65 & 0.85 & 0.30 \\
    TCV & 0.20 & 0.88 & 0.25 \\ 
    DIII-D & 0.67 & 1.67 & 0.50 \\
    NSTX & 0.58 & 0.85 & 0.23 \\
    NSTX-U & 0.62 & 0.85 & 0.30 
  \end{tabular}
  \caption{Parameters for several tokamaks used in simulations supporting an empirical scaling for $\chi_x$ (NSTX and NSTX-U parameters are shown for information but not simulated).}
  \label{tbl:device_conditions}
\end{table}

\begin{figure}
  \includegraphics[width=\linewidth]{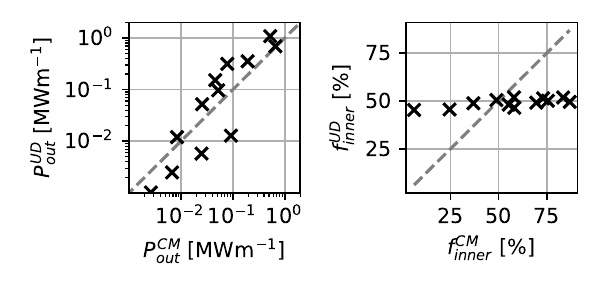}
  \caption{Comparison of $P_{out}$ (left) and $f_{inner}$ (right) by solving the CM physics (eqs. \ref{eq:orig_vort} - \ref{eq:orig_p}) and using an umbrella diffusive model (eqs. \ref{eq:orig_p}, \ref{eq:chi_mix} and \ref{eq:chi_x_fit_2}) across a range of SF divertor conditions (see table \ref{tbl:device_conditions}). The line $y=x$ is shown in grey on each plot.}
  \label{fig:ud_fit_test}
\end{figure}

We may use this fit to estimate $\chi_x$ in the umbrella diffusive (UD) model and compare predictions of $P_{out}$ to the turbulent calculation: this is shown in fig. \ref{fig:ud_fit_test}a. While this approach gives reasonable results for $P_{out}$ (root mean square error is 0.12 MWm$^{-1}$), it fails to capture the fractional heat deposited in each divertor leg. In fig. \ref{fig:distorted_psi} and table \ref{tbl:power_fracs}, we saw that the CM can result in a modification to the flux surfaces close to the X-points which redirects exhaust power in a way not captured by the UD model. This is seen most clearly in the fraction of exhaust power predicted to arrive on the inner divertor, $f_{inner}$, which is plotted for the UD and CM models in fig. \ref{fig:ud_fit_test}b: the UD model does not capture the variation in $f_{inner}$ at all. Clearly, based on the results shown in fig. \ref{fig:topo_change}, the picture is complicated further for larger $d_{xx}$.



An alternative approach to capturing the CM physics in a multi-fluid time-averaged transport model is to assume a pinch velocity in the null region: this better captures the actual CM dynamics at onset of convective clockwise motion bringing hot plasma from just above the X-points into the divertor (see fig. \ref{fig:churn_onset}). Working in polar $(r,\phi)$ coordinates in the poloidal plane centred at $x=y=0$, a simple model for this convection is 
\begin{equation}
  \vb{u}_\perp = -u_\perp \hat{\phi},
\end{equation}
where $u_\perp$ is a Gaussian profile over the convective zone,
\begin{equation}
  u_\perp = u_x e^{-\left(r/r_{cz}\right)^2}
\end{equation}
and $u_x$ is a parameter similar to $\chi_x$ in the UD model. This model can be made to reproduce both $P_{out}$ and $f_{inner}$ for the right choice of $u_x$. However, the CM dynamics results in clockwise rotation when $\pdv{p}{y}$ is positive (fig. \ref{fig:churn_onset}). Imposing this as a constraint in this pinch model, i.e. $u_x>0$, always results in $f_{inner} < f_{outer}$ as plasma fluid parcels travelling from just above the X-points deposit their energy in the outer legs first before reaching the inner legs. The equilibrium CM dynamics is a function of both this convective transport and the modification to flux surfaces in the null region, which is not captured by either of these simpler models. At this stage, we recommend eq. \ref{eq:chi_x_fit_2} for rough estimates of energy transport across the X-points in SF experiments, but a more accurate model which captures the directionality of the CM-induced transport requires further work. 

\section{Discussion}
\label{sec:discussion}



In SF experiments, an increase in power and particle fluxes to secondary strike points is evidence of enhanced transport across the null region. This effect has been shown in the CM simulations here in a SF-plus configuration in fig. \ref{fig:Q_legs}. However, power is only effectively shared across three legs at low $d_{xx}$ here, which was also shown in table \ref{tbl:power_fracs} and the related discussion. Note that this is similar to the results from TCV presented in \cite{reimerdes_power_2013} (see fig. 3 there). 


At any given $\beta_{pm}$, there may exist an optimal point in the $(d_{xx}, \theta)$ parameter space for even mixing across the divertor legs due to the CM, i.e. where $f_i \simeq 0.25$ for all $i$. Given that $f_3$ is small across all values of $d_{xx}$ and $\beta_{pm}$ shown in fig. \ref{fig:Q_legs}, and the fact $\theta$ does not influence the CM turbulence at low $d_{xx}$, we can conclude that the near-exact SF is unlikely to yield optimal mixing. This conclusion holds for the simulations at small $d_{xx}$ in conditions representative of TCV and DIII-D (see table \ref{tbl:device_conditions}), where $f_3$ does not exceed $11\%$.

Two further points are worth discussing regarding power sharing across divertor legs. Firstly, despite that fact the mixing is not optimal, the near-exact SF at high $\beta_{pm}$ (fig. \ref{fig:Q_legs}a, lower panel) has reduced the highest power fraction received by any leg from over 80\% at high $d_{xx}$ to under 50\% at small $d_{xx}$. Secondly, the CM-induced exhaust properties vary strongly with $\beta_{pm}$. 
This indicates that the turbulence during ELMs (high $\beta_{pm}$) is likely to redirect power differently to the situation between ELMs (low $\beta_{pm}$), above and beyond activating or deactivating the mode. Therefore, the ability for the CM to improve SF divertor performance should be assessed over several ELM cycles. This is highlighted by the results shown in figs. \ref{fig:topo_change} and \ref{fig:P2_P3}: in a hypothetical scenario in which leg 3 receives some inter-ELM heat flux and leg 2 receives significant during-ELM heat flux, the CM would facilitate redistribution in the time-integrated deposited energy to each target plate, but may not help to reduce peak heat loads.  

In SF experiments carried out to date, the churning mode has not been observed directly. Indirect evidence includes the increase in transport across the nulls as $\beta_{pm}$ is increased: predicted in \cite{ryutov_churning_2014}, this has been observed in \cite{soukhanovskii_developing_2018} and \cite{vijvers_power_2014} and is also observed in the simulations here (see fig. \ref{fig:nscan}). A detailed comparison to SF experiments is beyond the scope of this study, but the results in figs. \ref{fig:topo_change} and \ref{fig:P2_P3} indicate that an experiment could be designed to replicate the redirection of exhaust power from leg 3 to leg 2 in a SF experiment with $d_{xx} > \lambda_x$ and $\theta \sim 15^\circ$. 

It is important to note that the CM physics does not depend directly on the toroidal magnetic field. On the other hand, an alternative mechanism of cross-field transport in the SF divertor proposed in \cite{canal_enhanced_2015}, $\vb{E}\times \vb{B}$ drifts from strong parallel electric fields, would be sensistive to both its magnitude and direction. As such, an experiment involving a scan in toroidal field strength and direction (clockwise/anticlockwise) may provide strong evidence for or against the CM playing a significant role in SF divertor transport. Future work will involve a more detailed comparison of the predictions of the CM and the mechanism proposed in \cite{canal_enhanced_2015}.

The change in magnetic geometry shown in figs. \ref{fig:distorted_psi} and \ref{fig:topo_change} highlights a potential source of error in the magnetic equilibrium reconstruction of SF experiments: outputs from Grad-Shafranov solvers such as EFIT may not capture this distortion of the flux surfaces. In the simulations here we find that the power delivered to each divertor leg is highly sensitive to the SF parameters $\beta_{pm}$, $d_{xx}$ and $\theta$ (see fig. \ref{fig:f_inner_theta}), largely due to the change in geometry in the null region. This poses a challenge in transport modelling of such experiments, but as noted above it may also assist in determining the role of the CM experimentally. 


This study has focussed exclusively on the SF divertor. As discussed in \cite{ryutov_churning_2014}, the expected size of the convective zone in a conventional, single-null divertor is 
\begin{equation}
  r_{cz} \simeq 0.44 \beta_{pm} \varepsilon a,
\end{equation}
which results in $r_{cz}$ roughly an order of magnitude smaller than in a SF. However, during strong ELMs $r_{cz}\sim$ few cm is plausible, and therefore the CM may contribute to asymmetries in the inner/outer divertor heat flux as seen here. Therefore, further work exploring this in detail may be worthwhile. 

\section{Conclusions}
\label{sec:conclusions}

This study was motivated by a desire to better understand the CM physics in conditions relevant to experiments at MAST-U and other existing small- and medium-sized tokamaks. This has necessitated including the effect of field-aligned thermal conduction into a model of the CM, which allows capturing the competition between conductive and thermal transport in this regime. It has also required modelling realistic SF configurations with finite $d_{xx}$ and $\theta$, both of which are shown to strongly influence the CM dynamics.  
The model has been implemented in BOUT++ and used to simulate SF divertors across a range of the relevant parameter space. 


We have found that the CM can induce additional transport across the X-point region in the near-exact SF in MAST-U conditions for $\beta_{pm}\gtrsim 8$ \%. For a SF with large $d_{xx}$, this transport scales roughly linearly as $d_{xx}$ is reduced until reaching a saturation around $d_{xx}\simeq 2.5$ cm.  

The CM turbulence reaches quasi-equilibrium on hydrodynamic timescales, $t\sim $ few ms. After saturation, the position of the separatrices at the outer midplane can continue to fluctuate by 1-2 \% of the minor radius on timescales slightly longer than the predicted CM convective timescales in the null region, tens of $\mu$s. 

The CM results in a modification to the flux surfaces in the null region. This effect, combined with the convective transport, influences the exhaust physics in the simulations. Accordingly, the fractional exhaust power going to each divertor leg (or alternatively to the inner/outer divertor) is highly sensitive to  $\beta_{pm}$, $d_{xx}$ and $\theta$. The CM can also result in a change in magnetic topology in the null region when $\theta$ is small, $\theta\sim 15^\circ$. 

Across all simulations the fractional exhaust power going to the lower left divertor leg is small, $f_3<15$ \%, and it typically receives the lowest exhaust power of all legs in regimes where the CM turbulence is active. This has also been seen experimentally at TCV \cite{reimerdes_power_2013,vijvers_power_2014} and DIII-D \cite{soukhanovskii_developing_2018}, although the simulations here are for different divertor parameters.

A diffusive model of CM transport has been assessed and been shown to predict the total transport across the null region. But this model fails to predict the power deposited in each individual divertor leg due to the change in magnetic geometry induced by the CM. A convective pinch model, while better capturing the dynamics of the CM, also does not succesfully capture asymmetries in the CM transport. 

While a detailed experimental comparison is beyond the scope of this study, we have shown that the CM remains a plausible candidate for the additional transport across the X-points observed in SF experiments. Note in particular that the CM model of eqs. \ref{eq:orig_vort} - \ref{eq:orig_p} does not include any atomic processes or radiation. These effects would strongly influence exhaust properties and may be different for each divertor leg, and would be required for a comparison to SF experiments. 



\section{Acknowledgements}
This work was carried out under the auspices of the U.S. Department of Energy by Lawrence Livermore National Laboratory under Contract DE-
AC52-07NA27344. Simulations were carried out using resources of the National Energy Research Scientific Computing Center (NERSC), a U.S. Department of Energy Office of Science User Facility. We would also like to thank D.D. Ryutov for helpful and enlightening discussions during development of this work.

\appendix 
\section{Numerical implementation of field-aligned thermal conduction}
\label{app:cond}

We seek an accurate finite difference stencil for the term $\div{\vb{q}_\parallel}$, where $\vb{q}_\parallel = \kappa_\parallel \grad_\parallel{T_e}$. The unit vector parallel to the magnetic field, $\vb{b}$, has components in both toroidal and poloidal directions but we are only interested in the projection of $\vb{q}$ to the poloidal plane here. The parallel derivative is $\grad_\parallel = (\vb{b} \cdot \grad)\vb{b}$. As described in sec. \ref{sec:simulations}, a uniform, rectangular grid in the $x$ and $y$ directions defining the poloidal plane is used. Grid widths are $\Delta x$ and $\Delta y$. Note that the $\div{\vb{q}_\perp}$ term does not require special treatment, and a traditional finite difference stencil is used here. The subscript will be dropped from $T_e$ from hereon to lighten the notation.



The method presented \cite{gunter_modelling_2005}, which involves computing fluxes on cell corners, was implemented and found to be insufficiently accurate here. Alternative approaches such as those in \cite{stegmeir_field_2016,soler_new_2020} rely on one grid axis (e.g. the toroidal direction) being approximately field aligned: there is no such axis in the geometry of the problem studied here. Nevertheless, we develop a method based on those in \cite{stegmeir_field_2016,soler_new_2020}, which involve interpolating $T$ at points along the field lines to calculate $\grad_\parallel T$ combined with a support operator method (SOM) for $\div{\vb{q}_\parallel}$ \cite{shashkov_support-operator_1995}.  

\begin{figure}
  \includegraphics[width=\linewidth]{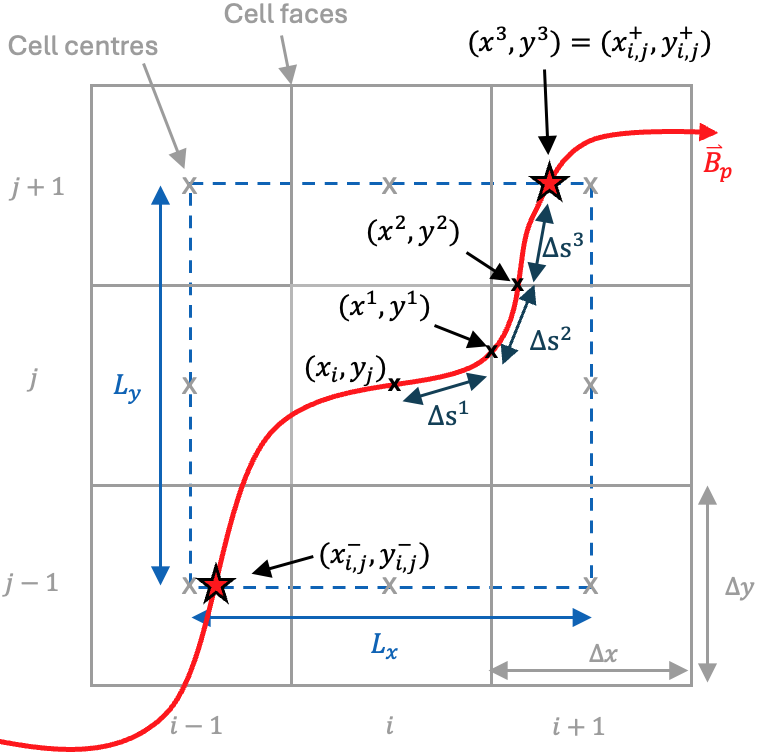}
  \caption{Geometry of the line tracing process to evaluate parallel derivatives for an arbitrary poloidal magnetic field $\vb{B}_p$. The bounding box is the region within the dashed blue lines. }
  \label{fig:linetrace}
\end{figure}

The unit vector $\vb{b}$ is
\begin{equation}
  \begin{aligned}
      \vb{b} &= \frac{1}{B} \left( B_x \hat{x} + B_y \hat{y} + B_z \hat{z} \right) \\
      & = b_x \hat{x} + b_y \hat{y} + b_z \hat{z}
  \end{aligned}
\end{equation}
where $\hat{z}$ points into the poloidal plane and $B$ is the magnitude of the magnetic field. Starting at point $(x,y)$, we can trace along the local magnetic field to another point on the poloidal plane $(x^+,y^+)$ via 
\begin{equation}
  x^+ = x + \int_0^{y^+}{\frac{b_x}{b_y}dy}
\end{equation}
\begin{equation}
  y^+ = y + \int_0^{x^+}{\frac{b_y}{b_x}dx}
\end{equation}
We now draw a box with sides $L_{x,y}$ around $(x,y)$ and discretise. The $x$ and $y$ cell indices are $i,j$, and we will also assume $\vb{b}$ is constant within the area covered by each cell (i.e. changing direction only at cell faces). Thus, starting at the coordinates of the centre of a given grid cell $(x_i,y_j)$, we have
\begin{equation}
  x_{i,j}^+ = x_i + \sum_{k=1}^{N_c}{\frac{b_x^k}{b_y^k} \Delta y^k},
\end{equation}
and
\begin{equation}
  y_{i,j}^+ = y_j + \sum_{k=1}^{N_c}{\frac{b_y^k}{b_x^k} \Delta x^k},
\end{equation}
where we trace along $\vb{b}$, intercepting cell faces until eventually reaching the bounding box. The intercept index $k$ runs from 1 to $N_c$, which is the total number of intercepts up to and including a collision with the bounding box. The coordinates of each intercept are $(x^k, y^k)$, so $\Delta x^k = x^k - x^{k-1}$ and $\Delta y^k = y^k - y^{k-1}$ where $(x^0,y^0)=(x_i,y_j)$. The vector $\vb{b}^k$ is evaluated at the centre of the cell containing the line segment from intercept $k-1$ to $k$. 

We define $T^+_{i,j}$ as the temperature at point $(x_{i,j}^+,y_{i,j}^+)$, calculated here via a linear interpolation from surrounding cell centres. To avoid gaps in the stencil which can introduce grid scale oscillations, we set $L_x=2\Delta x$ and $L_y=2\Delta y$. This ensures $T^+_{i,j}$ is evaluated using cells adjacent to $T_{i,j}$. 

The geometry of this problem is shown in fig. \ref{fig:linetrace}, where an example with $N_c=3$ is shown. 

The distance traversed along the total (poloidal plus toroidal) magnetic field in this line tracing process is 
\begin{equation}
  \Delta l^+_{i,j} \simeq \sum_{k=1}^{N_c} \Delta s^{k} \sqrt{\frac{(b_z^k)^2} {(b_x^k)^2 + (b_y^k)^2} + 1}
\end{equation}
where $\Delta s^k = \sqrt{(\Delta x^k)^2 + (\Delta y^k)^2}$. 
The parallel gradient is thus approximated as 
\begin{equation}
  \grad_\parallel T_{i,j} \simeq \frac{T_{i,j}^+ - T_{i,j}}{\Delta l^+_{i,j}} \equiv Q^+ (T_{i,j}),
\end{equation}
where the right hand side indicates this defines a linear operator, $Q^+$, acting on $T_{i,j}$. At a given cell $i,j$, this operator approximates $\grad_\parallel$ at the point half way between $(x_i,y_j)$ and $(x_{i,j}^+,y_{i,j}^+)$. 

We now carry out the same procedure in reverse, tracing along the direction anti-parallel to $\vb{b}$ to define an equivalent operator $Q^-$. The parallel diffusion operator, $\div{(\kappa_\parallel \grad_\parallel T)}$, is estimated using a similar SOM procedure as described in e.g. \cite{stegmeir_field_2016}. The result employs the transpose of $Q$, $[Q]^T$, and we take an average of the values calculated using $Q^+$ and $Q^-$. Thus we arrive at 
\begin{equation}
  \begin{aligned}
  \div{(\kappa_\parallel \grad_\parallel T)} \simeq \frac{1}{\Delta l^+ + \Delta l^-} \bigg( & \Delta l^- [Q^-(\kappa_\parallel Q^-(T))] ^T  \\
  &+ \Delta l^+ [Q^+(\kappa_\parallel Q^+(T))] ^T \bigg), 
  \end{aligned}
\end{equation}
where the $i,j$ indices on all discretised quantities have been omitted for clarity. 

We are assuming constant $\kappa_\parallel$ here, but spatially varying conductivity can be treated by evaluating $\kappa_\parallel^{\pm}$ at the point half way between $(x_i,y_j)$ and $(x_{i,j}^{\pm},y_{i,j}^{\pm})$. 

For a magnetic field varying on spatial scales larger than the grid spacing, this approach can be made more efficient without significant loss of accuracy by assuming $\vb{b}$ is constant within the bounding box around each cell. This allows skipping over the intermediate incercepts to find $(x_{i,j}^{+},y_{i,j}^{+})$. Further simplification arises from the fact $(x_{i,j}^{-},y_{i,j}^{-})=(x_i-x_{i,j}^+,y_j-y_{i,j}^+)$ and $\Delta l_{i,j}^+ = \Delta l_{i,j}^-$. To reduce the compute time involved in long-run CM simulations, this approximation was used in the results presented here. 

\bibliographystyle{unsrt}
\bibliography{zotero_lib.bib}

\end{document}